\documentclass{ws-mpla}

\begin{document}

\markboth{Upadhyay, Jain, Khemchandani, Kelkar}
{Pairwise final state interactions in the
$p\,d \, \rightarrow \, p \,d\,\eta$ reaction near threshold}

\catchline{}{}{}{}{}

\title{PAIRWISE FINAL STATE INTERACTIONS IN THE 
$p\,d \, \rightarrow \, p \,d\,\eta$ REACTION NEAR THRESHOLD}

\author{\footnotesize 
N. J. UPADHYAY\footnote{email: njupadhyay@gmail.com},  
B. K. JAIN\footnote{email: brajeshk@gmail.com}} 

\address{Department of Physics, University of Mumbai\\ 
Vidyanagari, Mumbai - 400 098, INDIA.}

\author{K. P. KHEMCHANDANI\footnote{email: kanchan@teor.fis.uc.pt}}

\address{Centro de F\'isica Computacional, 
Departamento de F\'isica\\
Universidade de Coimbra, P-3004-516 Coimbra, PORTUGAL.}

\author{N. G. KELKAR\footnote{email: nkelkar@uniandes.edu.co}}

\address{Departamento de F\'isica, Univsersidad de los Andes\\
Cra.1E, No.18A-10, Bogota, COLOMBIA.}

\maketitle

\pub{Received (Day Month Year)}{Revised (Day Month Year)}

\begin{abstract}
A model for the $p\,d\,\rightarrow\,p\,d\,\eta$ reaction
published earlier, including the final state interaction (FSI) of
all particles, is revisited to investigate the low energy data on 
this reaction. The three body problem of $p$-$d$-$\eta$ scattering in 
the final state is approximated in terms of pairwise interactions 
between the three particles in the final state. Apart from a 
comparison with some preliminary data, two new findings relevant to 
the near threshold data analysis are reported. The first one points
toward the limitations of an FSI factor used conventionally to
extract the eta-deuteron scattering length and infer subsequently on
the existence of eta-mesic states. The second result emphasizes the
role of the $p-d$ FSI and the strong Coulomb repulsion near threshold.
Finally, a comparison of the above model calculation with low energy
data, excludes very large eta-nucleon scattering lengths.

\end{abstract}

\ccode{PACS Nos.: 25.10.+s, 25.40.Ve, 24.10.Eq}

\section{$\eta$ meson production}	

The strong and attractive nature of the $\eta-N$ interaction in the
$s-$wave \cite{bhale} manifests itself in the sharp enhancement of 
cross sections in $\eta$-producing reactions near threshold 
\cite{reactdata1,reactdata2,reactdata3,reactdata4,reactdata5}. 
This enhancement was attributed to the possible 
existence of exotic $\eta-$nucleus quasi-bound states \cite{haili} 
first proposed by Q. Haider and L. C. Liu. The experimental findings 
motivated theoretical searches of metastable states of $\eta$-mesic 
nuclei (see Refs. 7-15 for
some theoretical predictions and an extensive list of references of
other theoretical searches), as well as experimental
searches for the evidence of such states \cite{pfieff}
in the eta meson producing reactions.

To investigate this possibility in the three nucleon system, in the 
recent past total as well as differential cross sections for the 
$p\,d\,\rightarrow\,p\,d\,\eta$ reaction have been measured \cite{bil}
at the CELSIUS storage ring of The Svedberg Laboratory, Uppsala, 
using the WASA/PROMICE experimental apparatus for excess energies 
($Q$) $\ge$ 14 MeV. Due to the strong energy dependence of the total 
cross section close to threshold, the data at low energies are however
better suited to study the effect of the final state interaction 
(FSI) and explore the existence of quasi-bound eta-nuclear states. 
For $Q\,<$ 14 MeV, two data sets exist for the 
$p\,d\,\rightarrow\,p\,d\,\eta$ reaction.
One measurement was performed with the SPESIII spectrometer
at SATURNE for two excess energies, namely, $Q\,=$ 1.1 MeV and 3.3 MeV
\cite{hib}. The other data set was obtained using the COSY-11 
detection system at the COSY-J\"ulich accelerator for $Q\,=$ 3.2 MeV, 
6.1 MeV and 9.2 MeV \cite{pis}. These data show enhancements of more 
than an order of magnitude in comparison to the theoretical plane wave
total cross sections, suggesting thereby a strong FSI effect.

\section{Interpreting the near threshold 
$p\,d \, \rightarrow \, p \,d\,\eta$ data}

At low energies, ignoring the $p-d$  FSI, an
averaged squared $\eta-d$ production amplitude, $|F(k)|^2$ (which is 
the ratio of the cross section for the production of the $\eta-d$ 
system to arbitrarily normalized phase space) is traditionally 
extracted from data in terms of the scattering length $A_{\eta\,d}$ 
by parametrizing \cite{bil} it as,
\begin{equation}\label{eq1}
|F(k)|^2\,=\,\frac{f_B} {|1\,-\,i\,k\,A_{\eta\,d}|^2} \, ,
\end{equation}
where, $f_B$ is considered a (basically unknown) scale factor and in
principle depends upon the particular reaction studied. At the low
energies considered here, this factor is practically energy 
independent. $k$ is the relative momentum of the $\eta-d$ system.
Within this parametrization, the analysis of the WASA data
(for $Q\,\ge$ 14 MeV), using different values of $A_{\eta\,d}$ taken 
from Ref. 20 suggests a preference for 
$A_{\eta\,d}\,=\,1.64 + i2.99$ fm (as mentioned in Ref. 17). This 
value of the $\eta-d$ scattering length is obtained in Ref. 20  
using an $\eta-N$ scattering length of, 
$a_{\eta\,N}\,=\,0.55 + i0.30$ fm.

Conclusions drawn about $A_{\eta\,d}$ and subsequent inferences made 
about the possible existence of quasi-bound $\eta-d$ states, from 
such studies, may suffer from limitations and uncertainties because 
of the reasons listed in the following two subsections.
\begin{figure}[th]
\centerline{\psfig{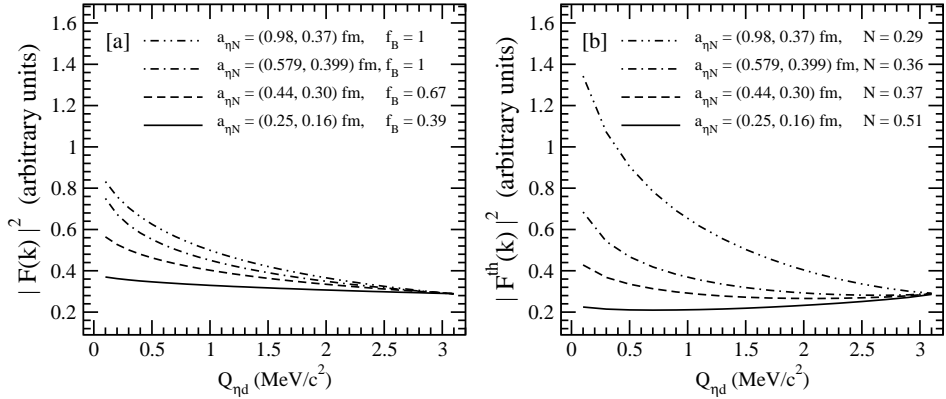}}
\vspace*{8pt}
\caption{Comparison of $|F(k)|^2$ obtained using:
(a) the factorized enhancement factor prescription
of $|F(k)|^2$ and (b) with the $|F^{\rm th}(k)|^2$
theoretically obtained from the full calculations generating the 
$\eta-d$ scattering wave function solving the Lippmann-Schwinger 
equation and using $T_{\eta\,d}$ constructed from $A_{\eta\,d}$. This 
calculation is done for $Q\,\sim$ 6 MeV.}
\end{figure}

\subsection{The scattering length description of $|F(k)|^2$ could be 
misleading}

The scattering length enhancement factor description involves the
factorization of the $p\,d\,\rightarrow\,p\,d\,\eta$ reaction 
amplitude into the $\eta$ production amplitude and an FSI factor.
This may not be correct. Fig. 1 exhibits this. We plot here 
$|F(k)|^2$  using, (i) the factorized enhancement factor prescription 
(Eq. (\ref{eq1})) and (ii) using the definition,
\begin{equation}
|F^{\rm th}(k)|^2\,=\,\frac{(d\sigma\,/\,dM_{\eta\,d})^{\rm \scriptstyle
{theoretical}}}{\rm phase\,\,space},
\end{equation}
where, $(d\sigma\,/\,dM_{\eta\,d})^{\rm \scriptstyle
{theoretical}}$, is evaluated in a theoretical
model involving the FSI of the $p$, $d$ and $\eta$ \cite{bkj1}.
Here the $\eta-d$ FSI is included by numerically solving the
Lippmann-Schwinger (LS) equation (as described in the next section).
The $t$-matrix for $\eta$-d elastic scattering,
$T_{\eta\,d}$, entering the LS equation is
constructed from $A_{\eta\,d}$ given in Ref. 20.
For the definition of $(d\sigma\,/\,dM_{\eta\,d})^{\rm \scriptstyle
{theoretical}}$ and the phase space details, see 
\footnote{For the definition of the cross section and the phase
space, see Eq. (1) and Eq. (25) of Ref.  21 and
Eq. (4) of Ref. 21 or Eq. (\ref{tfull}) of this article for the
$t$-matrix which has been used to calculate the  cross sections.}.

We plot the squared amplitude obtained from Eqs. (1) and (2) in 
Fig. 1(a) and 1(b) respectively, as a function of the excess energy, 
$Q_{\eta d}$, in the $\eta-d$ centre of mass system. In order to 
highlight the slopes of these results for different $a_{\eta\,N}$,
we choose values for the normalizing constant, $f_B$, in case of the
amplitude obtained from Eq. (1), such, that the $|F(k)|^2$ has the 
same value for the maximum value of $Q_{\eta d}$ ($\sim$ 3.1 MeV) 
shown in Fig. 1(a). To do the same for the results obtained from 
Eq. (2), we multiply the results by a normalizing constant, $N$.
We observe that, as we approach the threshold, the curves
corresponding to the full calculations for different values of
$A_{\eta\,d}$ (shown in Fig. 1(b)) open up much faster than the 
approximate calculations (Fig. 1(a)). This means that the value of 
$A_{\eta\,d}$ extracted by fitting the data by using Eq. (1) could be 
larger than that obtained by using the full calculation.

\begin{figure}[th]
\centerline{\psfig{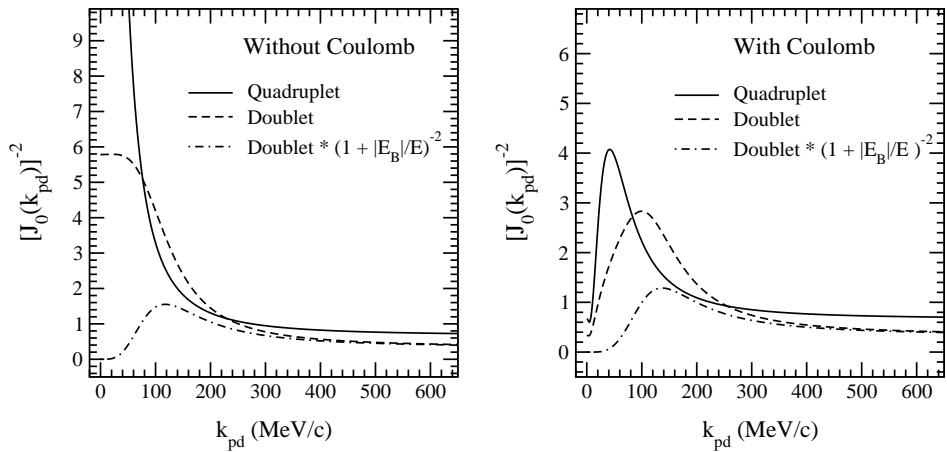}}
\vspace*{8pt}
\caption{Different contributions in the inverse of the $s-$wave
Jost function squared, $[J_0(k_{pd})]^{-2}$. The effect of including
Coulomb repulsion is also shown.}
\end{figure}

\subsection{$p-d$ FSI is stronger at low energies}

The $p\,d\,\rightarrow\,p\,d\,\eta$ reaction, in the final state, in
addition to the $\eta-d$ interaction, involves the $p-d$ interaction, 
which at low energies becomes important. There are two possible 
$s-$wave $p-d$ final states, the spin-doublet ($^2S_{1/2}$) and the 
spin-quartet ($^4S_{3/2}$) states. $^2S_{1/2}$ bears connection to 
the $p-d$ distribution in the $^3He$ bound state. Furthermore, since
both the proton and deuteron are positively charged, at low energies
there exists a sizable Coulomb repulsion between them too. The $p-d$ 
FSI in the present work is included through the multiplication
of an inverse Jost function \cite{bkj1}. This function is known to 
give the modification of the non-interacting wave function of two 
particles at short distances due to the interaction between them (see 
next section for the Jost function expressions or Ref. 21 for 
more details). In Fig. 2 we show the
inverse of the squared Jost function, $[J(k_{pd})]^{-2}$ for the
$s-$wave $p-d$ interaction. In order to show the effect of the 
Coulomb repulsion, we have plotted the Jost function for the doublet 
and the quadruplet states with and without inclusion of the Coulomb 
part separately. The contribution of the Jost function without 
Coulomb part can be obtained by keeping $C_o^2$ = 1 in 
Eq. (\ref{quad}). We also show the doublet Jost function multiplied 
by a factor $1\,+\,(|E_B| / E)$ to reflect the fact that the 
$^2S_{1/2}$ $p-d$ state has a bound state in ($^3$He) with a 
separation energy $|E_B|$ of 5.48 MeV. We see that at small values of 
$pd$ relative momentum the strong interaction in the $p-d$ system can 
enhance the wave function and hence the cross section, by a large 
amount. The Coulomb interaction, being repulsive reduces this effect. 
However, the overall effect is still an enhancement (up to a factor of
3 - 6).

The above discussion thus makes it clear that to learn reliably about
the $\eta-d$ interaction near threshold the data on the 
$p\,d\,\rightarrow\,p\,d\,\eta$ reaction in this energy range should 
be analyzed, as far as possible, with the full theoretical framework 
available and should include all the interactions. In what follows,
we have made an attempt to explain the new (preliminary) data 
available from Ref. 19 by using a formalism \cite{bkj1}
which includes the interaction between all the pairs in the final 
state.

In the following section we describe the formalism only briefly and 
move on to discussing results and conclusions. For more details of the
formalism we refer the reader to our previous articles 
\cite{bkj1,bkj2,bkj3}.

\section{$\eta$ Production and $\eta-d$ $t$-Matrices}

The cross sections for the $p \,d\, \rightarrow \, p \, d\, \eta$
reaction close to threshold are evaluated using the two step model 
for $\eta$ production, where the proton interacts with a proton (or 
neutron) in the deuteron in the first step.
This produces a deuteron and a $\pi^+$ (or $\pi^0$). The pion in the
second step interacts with another nucleon in the original deuteron 
and produces an $\eta$. The two steps thus are 
$p\,p\,(n)\,\rightarrow\,d\,\pi^+\,(\pi^0)$ and 
$\pi\,N\,\rightarrow\,\eta\,N$. The $t$-matrix for this model in the 
plane wave approximation is written as
\begin{eqnarray}
\nonumber
\langle\,|\,T_{p\,d\,\rightarrow\,p\,d\,\eta}\,|\,\rangle&=& \frac{3}{2}
\,i\,\sum_{m's}\int \frac{d\vec{P}}{2\,\pi)^3}\,\langle\,p\,n\,|\,d\,
\rangle\,\langle\,|\,T_{p\,p\,(n)\,\rightarrow\,d\,\pi^+\,(\pi^0)}\,|\,
\rangle
\\
&&\times\,\frac{1}{k_{\pi}^2\,-\,m_{\pi}^2\,+\,i\,\epsilon}\,\langle\,|\,
T_{\pi\,N\,\rightarrow\,\eta\,p}\,|\,\rangle,
\label{pwtm}
\end{eqnarray}
where $k_{\pi}$ is the four momentum of the intermediate pion.
The summation is over internal spin projections and the matrix element
$\langle\,p\,n\,|\,d\,\rangle$ represents the deuteron wave function
in momentum space for which we use the Paris parametrization
\cite{paris}.

The above kind of model for the production amplitude which involves a
two step process (and hence a three body mechanism for production) has
often been used for meson producing reactions at intermediate and low
energies. The importance of such a model for the $p \, d\, \to \,^3$He
$\eta$ reaction was first shown by Laget \cite{laget} and then tested 
with more recent data in Ref. 22 and 23. It has also been used for other 
meson producing reactions in Ref. 26 and 27. Though one worries 
about unitarity bounds on cross sections involving two step processes 
at high energies ($\gamma p$ scattering for example Refs. 28-30), 
such worries are not relevant for the present work. The model 
calculations done here are very close to threshold and in addition, 
the cross section for $\eta$ production saturates at high energies.

The pion production in the first step being at high energy, is 
described by an on-shell $p\,N\,\rightarrow\,d\,\pi$ $t-$matrix, 
which is parametrized using the data on this reaction \cite{arn}.
The $\langle\,|\,T_{p\,p\,(n)\,\rightarrow\,d\,\pi^+\,(\pi^0)}\,|\,
\rangle$ in Eq. (\ref{pwtm}) has been related to the amplitudes 
$\langle\,|\,f |\,\rangle$ given in Ref. 31 as
\begin{equation}
\langle\,|\,T_{p\,p\,(n)\,\rightarrow\,d\,\pi^+\,(\pi^0)}\,|\,\rangle 
= \sqrt{\frac{2 \pi^2 s_{p p\,\rightarrow\,d \pi}}{m_p^2 m_d p_f p_i}}
\times \langle\,|\,f |\,\rangle
\end{equation}
and we relate it to the cross section using
\begin{equation}
\frac{d\sigma}{d\Omega}\,=\,\frac{m_p^2\,m_d}{8\,\pi^2\,
s_{p p\,\rightarrow\,d \pi}}\,\frac{|\vec{p_f}|}{|\vec{p_i}|}\,
\frac{1}{4}\,\sum\limits_{spin\, projections} \langle\,|\,T_{p\,p\,(n)
\,\rightarrow\,d\,\pi^+\,(\pi^0)}\,|\,\rangle
\end{equation}
where $\vec{p_i}$ ($\vec{p_f}$) is the initial (final) momentum in 
the $p p\,\rightarrow\,d \pi$ centre of mass system and 
$\sqrt{s_{p p\,\rightarrow\,d \pi}}$ is the total energy of the same.

The second step $\pi\,N\,\rightarrow\,\eta\,N$ being near threshold, 
is described by an off-shell $t-$matrix which is taken from a coupled 
channel calculation \cite{bhale} which reproduces the 
$\pi\,N\,\rightarrow\,\eta\,N$ data well. We refer the reader to 
Ref. 32 for some recent interesting investigations of the
$\pi\,N\,\rightarrow\,\eta\,N$ data.

To take the FSI into account  we include the interactions between the
$p-d$ and $\eta-d$ pairs. The $\eta p$ interaction is implicit in
our description of the production vertex, which uses the $t-$matrix 
for $\pi\,N\,\rightarrow\,\eta\,N$. Though, in principle, a
finer treatment of the three body problem (involving the 3-body
$p$-$d$-$\eta$ scattering) would be more appropriate, we expect the 
present approach to work reasonably well for the problem under 
consideration. Some remarks in connection with this issue are in 
order here.

The choice of any approximation is guided by the characteristics of
the problem at hand. In our case, because of the large momentum
transfer ($\sim$ 770 MeV/c), the reaction takes place over a small 
volume of space. In such situations (first introduced by Watson and 
Migdal and subsequently discussed in most of the scattering theory
text-books \cite{tbooks1,tbooks2}),
an essential  feature of the cross sections is that the energy
dependence of the cross section due to FSI, factors out from that of
the primary production amplitude. The FSI factor, which has the 
correct high energy behaviour is usually taken to be the inverse of 
the Jost function. In the present paper we have used this 
prescription for the p-d interaction. Having done that, we could 
include the eta-d interaction fully, hence we did not resort to a 
similar factorization for it.
We included the eta-d interaction effect fully by keeping it inside 
the production amplitude and using an exact scattering wave function.
The above factorization method has been used in literature to describe
the low momentum behaviour of the meson 
\cite{moalem1,moalem2} and 
pion spectra \cite{dubach,bernard} measured in proton-proton collisions, 
and study the total cross sections in the 
$p \,p \,\to \,p \,K^+ \,\Lambda$ reaction \cite{rshyam}.

The transition matrix for the reaction, $ p\,d\,\to\,p\,d\,\eta$,
which includes the interaction between the $\eta$ meson and the 
deuteron is written as,
\begin{equation}\label{fulltmatrix}
T \,=\, \langle\,
\Psi^-_{k_{\eta\,d}}, \, \vec{k}_{p}^{\prime}\,; m_p^{\prime},
m_d^{\prime}\, |\, T_{p\,d\,\rightarrow\,p\,d\,\eta}\,|\, \vec{k}_p,
\vec{k}_d; m_p, m_d\, \rangle
\end{equation}
where $\vec{k_p}$ and $\vec{k_{p^\prime}}$ are the proton momenta in 
the initial and final states, respectively. $\vec{k_{\eta\,d}}$ is 
the $\eta$ momentum in the $\eta-d$ centre-of-mass. 
$m_p\,,\,m_d\,,\,m_{p^\prime}$ and $m_{d^\prime}$ are the spin 
projections for the proton and the deuteron in the initial and final 
states. The final state $\eta$-d wave function, 
$\Psi^{-*}_{k_{\eta\,d}}$ consists of a plane wave and a scattered 
wave and satisfies the Lippmann Schwinger equation for $\eta-d$ 
elastic scattering:
\begin{eqnarray}
\langle\, \Psi^-_{k_{\eta\,d}} \,| = \langle\,\vec{k_{\eta\,d}}\,| +
\int \frac{d\vec{q}}{(2\pi)^3}\,
\frac{\langle\,\vec{k_{\eta\,d}}\,|\,T_{\eta\,d}\,|\,\vec{q}\,\rangle}
{E(k_{\eta\,d})\,-\,E(q)\,+\,i\,\epsilon}
\,\langle\,\vec{q}\,|.
\end{eqnarray}
Here, $T_{\eta\,d}$ is the $t$-matrix for $\eta$-d elastic scattering.
Replacing the above equation in the transition matrix for the
$p\,d\,\rightarrow\,p\,d\,\eta$ reaction (Eq. (\ref{fulltmatrix})), 
leads to
\begin{eqnarray}
\nonumber
T&=&\langle\,\vec{k_{\eta\,d}}\,,\,\vec{k_{p^\prime}}\,;\,m_{p^\prime}
\,,\,m_{d^\prime}\,|\,T_{p\,d\,\rightarrow\,p\,d\,\eta}\,|\,\vec{k_p}
\,,\vec{k_d}\,;\,m_p\,,\,m_d\,\rangle
\\
\nonumber
&&+\,\sum_{m_{2^\prime}}\int \frac{d\vec{q}}{(2\pi)^3}\,\frac{\langle\,
\vec{k_{\eta\,d}}\,;\,m_{d^\prime}\,|\,T_{\eta\,d}\,|\,\vec{q}\,;\,
m_{2^\prime}\,
\rangle}{E(k_{\eta\,d})\,-\,E(q)\,+\,i\,\epsilon}\,
\\
&&\times
\langle\,\vec{q}\,,\,
\vec{k_{p^\prime}}\,;\,m_{2^\prime}\,,\,m_{p^\prime}\,|
\,T_{p\,d\,\rightarrow\,p\,d\,\eta}\,|\,\vec{k_p}\,,\vec{k_d}\,
;\,m_p\,,\,m_d\,\rangle\, ,
\label{tfull}
\end{eqnarray}
where the first term represents the $\eta$ production without any 
interaction with the deuteron and the second one takes care of
the rescattering of the $\eta$-d pair to all orders. It is indeed the
half-off-shell $t$-matrix, $T_{\eta\,d} $, which is reponsible for
converting the off-shell eta mesons in
the intermediate state to the on-shell ones in the final state.

Since the low energy $\eta N$ interaction is
dominated by the $S_{11}$ resonance N$^*$(1535), we perform a partial
wave expansion of the $\eta$-d $t$-matrix,
$T_{\eta\,d}(\vec{k_{\eta\,d}}, \vec{q})$
($\equiv \langle\,
\vec{k_{\eta\,d}}\,;\,m_{d^\prime}\,|\,T_{\eta\,d}\,|\,\vec{q}\,;\,
m_{2^\prime}\,\rangle$)
and retain only $s$-waves.
$T_{\eta\,d} (\vec{k_{\eta\,d}}, \vec{q})$
is half off-shell and we write it as a
product of an on-shell $t-$matrix and an off-shell
(which is actually half off-shell here) form factor.
Thus, we write
$T_{\eta\,d} (\vec{k_{\eta\,d}}, \vec{q})$ using $s$-waves only as,
\begin{equation}
T_{\eta\,d} (k_{\eta\,d}, q) \,
= \, T^{on}_{\eta\,d} (k_{\eta\,d}) \,
 \times \, g(k_{\eta\,d},\,q),
\end{equation}
where the $T^{on}_{\eta\,d} (k_{\eta\,d})$
is written in terms of the $\eta\,d$ scattering length as
\begin{equation}
T^{on}_{\eta\,d} = -\frac{2\pi}{\mu_{\eta d}} \times
\biggr({\frac{1}{A_{\eta\,d}}}\,-\,i\,k_{\eta\,d}\biggr)^{-1}.
\label{ton}
\end{equation}
We choose to define the off-shell form factor in terms of the
deuteron form factor as in Ref. 40,
\begin{equation}\label{deutform}
g(k_{\eta\, d},,\,q) = \,\int d\vec{r}\,j_0(rk_{\eta\,d}/2)\,|\phi_d(r)|^2\,
j_0(r\,q/2),
\end{equation}
where an average over the directions of the momenta has been performed
in order to extract the $s$-wave contribution. The deuteron wave 
function, $\phi_d(r)$, is written using the Paris parametrization 
\cite{paris}.

The $\eta\,d$ scattering length, used in Eq. (\ref{ton}), is taken 
from Ref. 20, where it has been deduced by solving three-body 
Faddeev type equations. This calculation includes the scattering of 
the $\eta$ on the break-up continuum states as well as on the deuteron
in the intermediate state. We use the $A_{\eta\,d}$ in this
work calculated for a wide range of $\eta-N$ scattering lengths since
the $\eta-N$ interaction is still not very well known.

For the $p-d$ interaction, as mentioned above, we use the Jost 
function, which is justified because the $\eta$ production involves 
high momentum transfer. Near threshold this value is around 700 MeV/c.
This means that the reaction occurs in a small interaction volume. 
The Jost function is given as in Ref. 21,
\begin{equation}
[J_o\,(k_{pd})]^{-2}\,=\,[J_o\,(k_{pd})]^{-2}_Q\,+\,
[(1\,+\,\frac{|E_B|}{E})J_o\,(k_{pd})]^{-2}_D .
\label{pdj}
\end{equation}
The expressions for spin quadruplet
(Q) and doublet (D) $[J_o\,(k_{pd})]^{-2}$ are given by
\begin{equation}
[J_o\,(k_{pd})]^{-2}_X\,=\, \frac{(k_{pd}^2\,+\,\alpha^2)^2\,(b^c_X)^2}{4}\,
\times \frac{1}{3\,C_o^2\,k_{pd}^2}\,\left[\frac{A_X}{
1\,+\,{\rm cot}^2\,\delta_X}\,\right]
\label{quad}
\end{equation}
where $X$ is either $Q$ or $D$, $A_D=1$ and $A_Q=2$. The factor 
$C_o^2$ above originates from the Coulomb interaction. Details 
regarding the parameters used in Eq. (\ref{quad}) can be found in 
Ref. 21.

The final amplitude is calculated by multiplying Eq. (\ref{pdj}) to 
the squared $t$-matrix obtained after including the $\eta-d$ 
interaction, i.e., Eq. (\ref{tfull}).

\section{$p d$ and $\eta d$ FSI effects on the total cross sections}

We present the total cross sections calculated, using our full 
amplitude described above, along with the data for excess energies up 
to 20 MeV in Fig. 3. The COSY-11 data shown in Fig. 3, 
though preliminary, seems to be in reasonable agreement with the 
previous data near threshold. In this figure we show separately the 
relative contribution of the different pieces in the FSI. The $\eta-d$
FSI has been calculated using an $\eta-d$ scattering length 
corresponding to $a_{\eta\,N}\,=\,0.98 + i0.37$ fm.
We find that (i) both, the $\eta -d$ and $p-d$ FSI affect the results
equally strongly and in the whole region of the
excess energies. (ii) Comparison of results with and without Coulomb
repulsion show that the Coulomb interaction reduces the cross section
by a large amount. In fact, it is because of this interaction
that the calculated cross sections come in accord with the 
experiments. This is one of the main result of our work.
\begin{figure}[th]
\centerline{\psfig{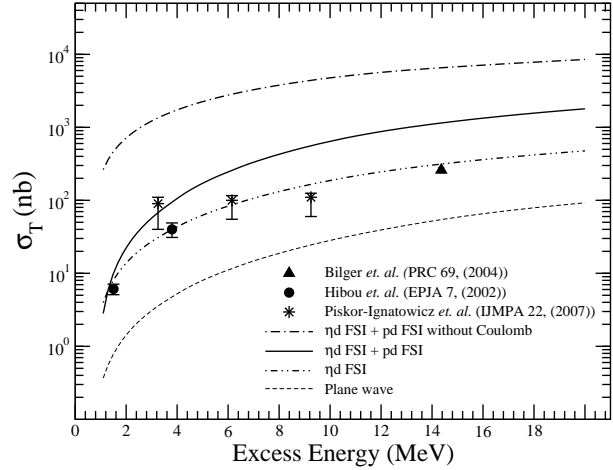}}
\vspace*{8pt}
\caption{Relative contribution from different pieces in the FSI. The 
results are for $a_{\eta\,N}\,=\,0.98 + i0.37$ fm.}
\end{figure}

We have also studied the sensitivity of the results to the $\eta-N$
scattering length which is still not so well known.
The results shown in Fig. 4(a) have been obtained using
the Paris parametrization for the deuteron form factor and four
different values of the $\eta-d$ scattering
lengths, $A_{\eta\,d}$. These $A_{\eta\,d}$ are
calculated for different input values of $a_{\eta N}$,
ranging from a weak to a strong $\eta-N$ interaction.
For the results presented here, these $a_{\eta N}$ values are 
$0.25 + i0.16$, $0.44 + i.30$, $0.579 + i0.399$, and $0.98 + i0.37$ 
fm. The different $A_{\eta\,d}$ results in the figure are identified 
by their corresponding $a_{\eta N}$ values. We see that (i) all the 
calculated cross sections show a strong FSI effect, and (ii) the 
results for $a_{\eta N}=0.44 + i.30$ and $0.579 + i0.399$ fm seem to 
pass through the maximum number of data points.
\begin{figure}[th]
\centerline{\psfig{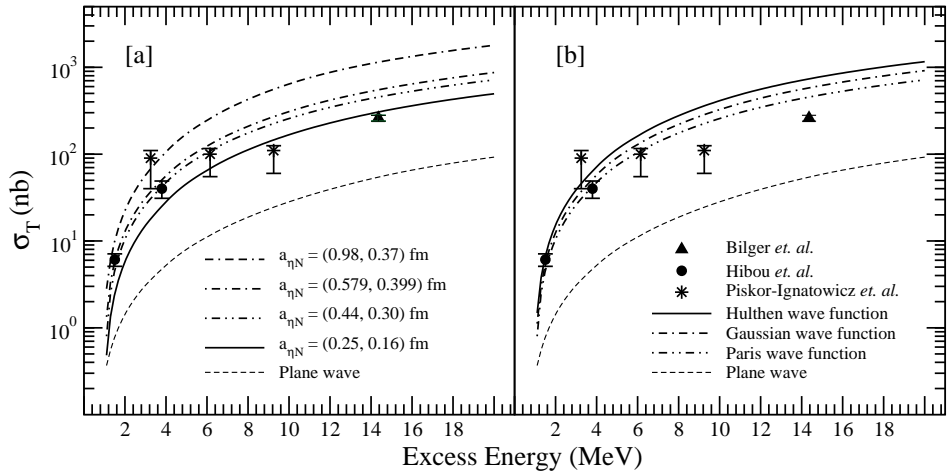}}
\vspace*{8pt}
\caption{A comparison of data \protect\cite{bil,hib,pis} with the 
total cross sections calculated including all the final state 
interactions, (a) using the Paris parametrized wave function for 
the deuteron in $\eta-d$ FSI and varying the values of $a_{\eta\,N}$  
and (b) for $a_{\eta\,N} = 0.44 + i.30$ but using different deuteron 
form factors in the FSI.}
\end{figure}

For the sake of completeness, in Fig. 4(b), we show the results 
obtained by using in Eq. (\ref{deutform}), prescriptions of the 
deuteron wave function other than the Paris form, which are also 
sometimes used in literature. We have thus calculated the total cross 
sections using the wave functions obtained from the Paris potential,
Hulthen potential, and the Gaussian form. The scattering length
$a_{\eta N}$ for the calculation in Fig. 4(b) has been kept fixed at 
$0.44 + i.30$ fm. The maximum difference in the total cross sections,
obtained due to the use of different wave functions, is found to be
a factor of the order of 1.6 at higher excess energies.

\section{Summary}

We have presented a theoretical analysis of the data on the
$p\,d\,\rightarrow\,p\,d\,\eta$ reaction near threshold using a model 
which was reasonably successful in reproducing the data at higher 
energies. We start by comparing the model calculations with a form of
the scattering amplitude often used for data analysis and investigate
the limitations of this form. The cross sections for the 
$p\,d\,\rightarrow\,p\,d\,\eta$ reaction at low energies (up to 
$\sim$ 20 MeV above threshold) are then evaluated within a two step 
model for $\eta$ production to investigate the importance of including
the FSI between all particles at low energies. The conclusions of 
this study can be summarized as follows:
\begin{enumerate}
\item{
Quantitative reproduction of the experimentally observed enhancement
in the cross section near threshold,
requires the inclusion of the interaction between
all three particles, namely, the proton, deuteron and the eta meson
in the final state. Both the strong and Coulomb $p-d$ FSI are found
to be important at low energies.}
\item{
The best agreement with experimental data seems to come from
the $\eta$-nucleon scattering length values of
$a_{\eta\,N} = 0.44 + i.30$ fm and $0.579 + i0.399$ fm.
Calculations based on the exact Alt-Grassberger-Sandhas equations
in Ref. 20 lead to $\eta - d$ scattering lengths of
($1.15\,+\,i 1.89$) fm and ($0.34\,+\,i 3.31$) fm
for the two above $a_{\eta N}$'s respectively.
}
\item{
In the energy range discussed here,
the scattering length description of the $\eta - d$ $t$-matrix,
$T_{\eta\,d}$, which enters the Lippmann-Schwinger equation
for the $\eta-d$ elastic scattering wave function,
seems to be sufficient.}
\item{The extracted value of the $\eta$-deuteron scattering length,
$A_{\eta\,d}$, from experimental $|F(k)|^2$ using the simplified
relation in Eq. (1) can be misleading.}
\end{enumerate}

\section*{Acknowledgments}

One of the authors, K. P. K., would like to thank the
Funda\c{c}\~{a}o para a Ci\^{e}ncia e a Tecnologia of
the Minist\'{e}rio da Ci\^{e}ncia,
Tecnologia e Ensino Superior of Portugal for financial support under
the contract SFRH/BPD/40309/2007.

\end{document}